\documentclass{ws-p8-50x6-00mod}

\begin{document}
\newcommand{\plumin}[2]{^{+#1}_{-#2}}
\newcommand{\plumi}[2]{\matrix{+#1\\[-3pt]-#2}}
\newcommand{\stth}{\sin^2 2\theta}
\newcommand{\cth}{\cos^2 \theta}
\newcommand{\sth}{\sin^2 \theta}
\newcommand{\csth}{\cos \theta\sin \theta}
\newcommand{\dm}{\Delta m^2}
\newcommand{\tsun}{\theta_{\mbox{\tiny Sun}}}
\newcommand{\csun}{\cos \tsun}
\newcommand{\phiosc}{\phi^{\mbox{\tiny osc}}_{i,z}}
\newcommand{\phissm}{\phi^{\mbox{\tiny SSM}}_i}
\newcommand{\delcor}{\delta_{\mbox{\tiny corr}}}

\title{A Unique Oscillation Solution to
the Solar Neutrino Problem?}

\author{M. B. Smy}

\address{Department of Physics and Astronomy\\
4182 Frederick Reines Hall,
University of California, Irvine, CA 92697-4575}


\maketitle

\abstracts{
A global two-neutrino oscillation
fit combining Super-Kamiokande solar neutrino
data with the solar neutrino rates measured by
Homestake, Gallex/GNO, SAGE and SNO
prefers a single allowed area, the Large Mixing Angle
solution, at about 95\% confidence level. The mass$^2$ difference
between the two mass eigenstates is 
$\Delta m^2\approx$3--25$\times10^{-5}$eV$^2$, the mixing angle
$\theta$ is $\tan^2\theta\approx$0.21--0.67.}

\section{Introduction}

Neutrino flavor oscillations are currently the most favored way
to understand the larger solar $^8$B neutrino interaction rate
measured by Super-Kamiokande\cite{sk} (SK) with respect to the
$^8$B rate reported by the SNO-collaboration\cite{sno}.
The SNO rate includes only $e$-type $^8$B neutrinos while
SK measures solar neutrinos via neutrino-electron elastic scattering
which has a small sensitivity to other (active) flavors as well.
The SK excess is then interpreted as a hint for appearance of other
active flavors in the $^8$B neutrino flux (born in the sun as
purely $e$-type).
Neutrino oscillations also explain the deficit of both rates as well as
the deficits of the ``Chlorine rate'' (Homestake\cite{cl}) and the ``Gallium rates''
(Gallex/GNO\cite{gallex}, SAGE\cite{sage}) with respect to the Standard
Solar Model\cite{ssm} (SSM). Chlorine and Gallium rates include
only $e$-type solar neutrinos.

The large neutrino flavor mixing
between the second and third generation inferred from atmospheric
neutrino data\cite{skatm} and the absence of an oscillation
signal in the CHOOZ reactor neutrino experiment\cite{chooz}
requires the neutrino flavor mixing between the first and third
generation to be small. Solar neutrino oscillations can therefore
be approximated by a two-neutrino description with the parameters
$\theta$ (mixing angle) and $\Delta m^2$ (difference in mass$^2$).
The mixing angle ranges between 0 and $\pi/2$, since
$\Delta m^2$ is defined to be positive.
For $\Delta m^2$ between $\approx 10^{-8}$eV$^2$ and
$\approx 10^{-3}$eV$^2$, the matter density in the sun and earth can
strongly affect the oscillation probability. For $\theta<\pi/4$,
resonant enhancement of the oscillations (MSW effect\cite{msw})
can occur in the sun. On the ``dark side'' of the parameter space
($\theta>\pi/4$), an anti-resonance can suppress the oscillations.
Below $\approx10^{-9}$eV$^2$ (quasi-vacuum/vacuum region),
the oscillation probability
is more affected by the oscillation phase than by matter effects.

\section{Parameter Estimation for Two-Neutrino Oscillations}

\begin{figure}[hbt]
\noindent a) \hspace*{13pc} b)
\vspace*{-1.5pc}

\epsfxsize=14.5pc
\epsfbox{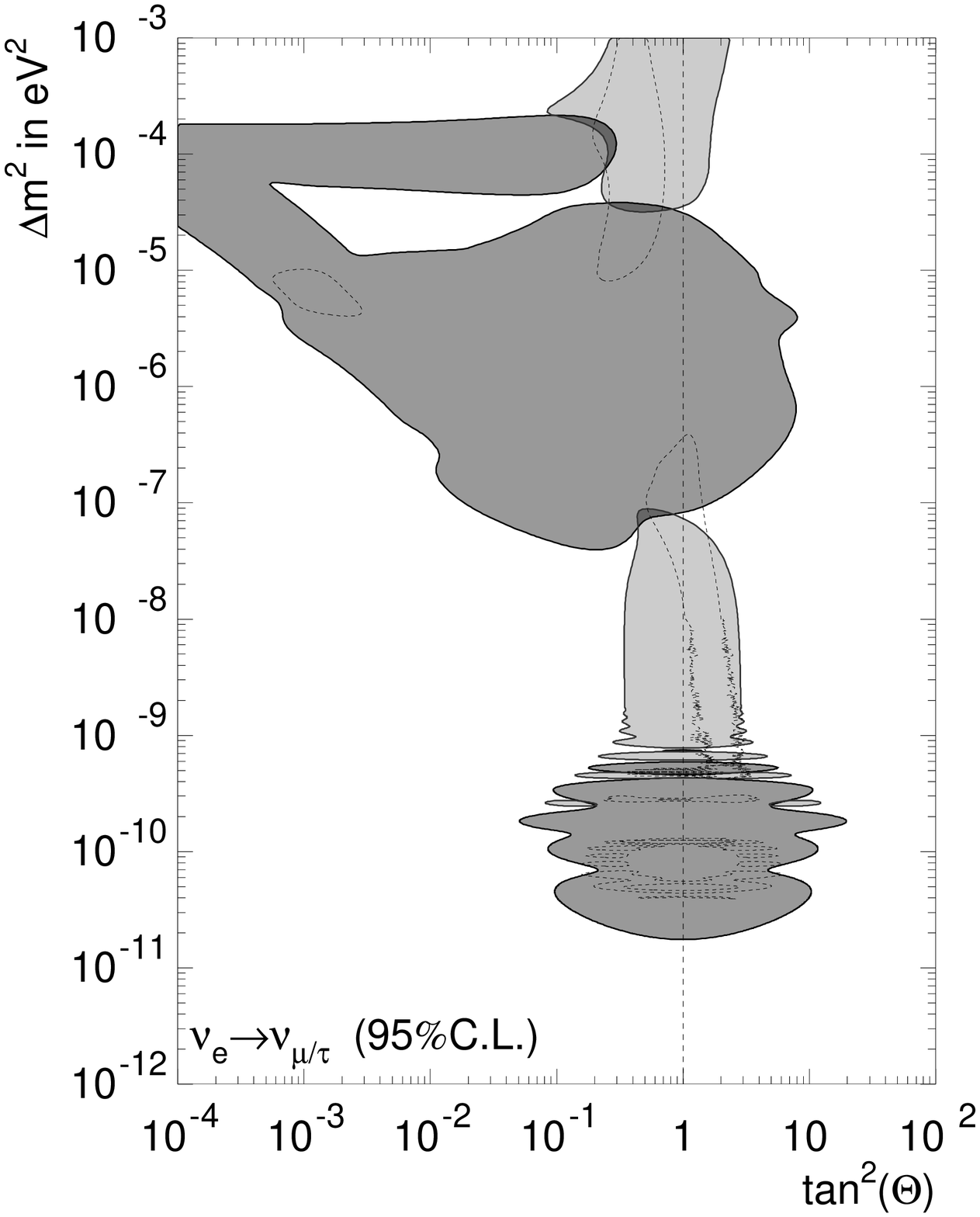} 
\epsfxsize=14.5pc
\epsfbox{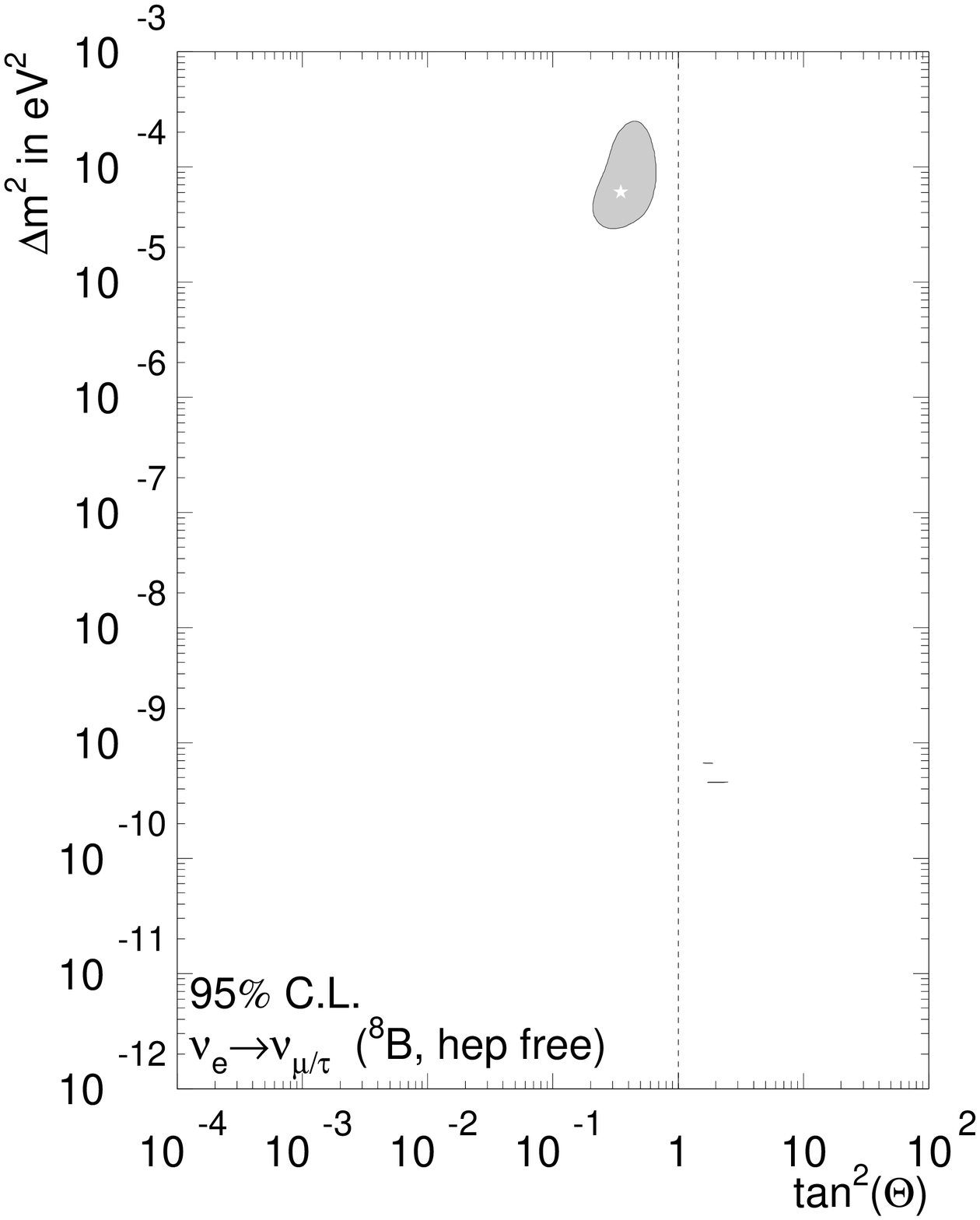} 
\caption{
SK Zenith Spectrum/SNO (a) and global fit (b)
contours at 95\% C.L. The dark-grey area in a) is
excluded by the shape of the SK Zenith Spectrum.
Adding the SNO rate measurement results in the 
(light-grey) allowed areas. Both regions do not depend
on neutrino flux predictions by the SSM. 
Overlaid are the allowed regions (inside dotted lines)
based on the Gallex/GNO, SAGE, Homestake and SNO rates
and the SSM.
In b) the
allowed areas from a global fit (adding the rates
of Gallex/GNO, SAGE and Homestake to the
SK Zenith Spectrum--SNO fit)
are predominantly
LMA solutions.}
\label{fig:global}
\end{figure}

The simplest way to constrain the oscillation parameters is
a $\chi^2$ fit to all $e$-type neutrino rates, that is
Gallex/GNO and SAGE (combined into a single ``Gallium'' number),
Homestake and SNO. The dotted lines of
figure~\ref{fig:global}a) show the allowed regions at 95\% C.L.
(which can be understood as an ``overlay''
of the allowed regions\cite{moriond} from the three experimental rates
mentioned above and their SSM-based predictions).
The area above $\Delta m^2\approx 10^{-5}$eV$^2$ 
near maximal mixing is called the Large Mixing Angle
(LMA) solution. The Small Mixing Angle Solution (SMA) is located
between $\Delta m^2\approx 10^{-6}$eV$^2$ and 
$\approx 10^{-5}$eV$^2$
at $\tan^2\theta\approx10^{-3}$.
The LOW solution is the large $\Delta m^2$ part of the
extensive
region(s) between $\approx 10^{-7}$eV$^2$ and $\approx 10^{-9}$eV$^2$.
The lower part of this region
is called the quasi-vacuum (quasi-VAC) solution.
The vacuum solutions (VAC) are below $\approx 10^{-10}$eV$^2$.
All regions have similar $\chi^2$
(LOW fits slightly worse).

To break the degeneracy in $\chi^2$ of these regions,
more information is needed than provided by the
suppression of the $e-$type rates.
Indeed, neutrino oscillations not only cause a
`disappearance' of the $e-$type fluxes but also
produce other 
neutrino flavors (appearance). Furthermore, they can induce distortions
of the neutrino spectrum and time variations of the
solar neutrino fluxes. In the MSW region, the time variations
arise from matter effects inside the earth (daily variations).
In the vacuum regions, the time variations are the consequence
of the change of the oscillation phase due to the yearly
variation of the oscillation baseline (distance between
sun and earth). 

SK can search for spectral distortions
through analysis of the spectrum of the recoiling electron
as well as daily or yearly time variations. The analysis of the
shape of the SK ``zenith angle spectrum''\cite{moriond,osc}
combines spectrum
and daily variation analyses. Neither spectral distortion
nor daily variation was found. 
The centerpiece of this work is
the preliminary SK zenith angle spectrum based on the entire SK
data set (1496 days). The numerical results are listed in
tables~\ref{tab:zenspec} and \ref{tab:zenspec2}.
Figure~\ref{fig:global}a) shows
the excluded regions (dark-grey) from the SK zenith angle spectrum
(for an explanation of the fit method see Appendix, eq.~\ref{eq:sk}).
The SMA and VAC regions are excluded since they require a
distorted $^8$B neutrino spectrum\cite{moriond}. The lower part of the
LMA and the upper part of the LOW solution 
predict daily variations\cite{moriond} and
are therefore disfavored.
The excluded areas
are independent of the SSM neutrino fluxes. The SK zenith
angle spectrum shape is best described by quasi-VAC
solutions.

\begin{table}[bt]
\caption{SK Rate and Uncertainty for eight energy bins.
The rates, statistical and systematic uncertainties (of the
spectrum shape) in the third column
are presented in units of SSM expectation. These systematic uncertainties
are assumed to be uncorrelated in energy.
The uncertainties in the fourth (uncertainty of the
$^8$B neutrino spectrum), fifth (uncertainty of the energy scale of 0.64\%)
and sixth column (uncertainty of the energy resolution of 2.5\%) are
fully correlated in energy (but uncorrelated with each other).
The combined uncertainty (last row) is
based on the sum of all bins of the zenith angle spectrum.
The combined rate has an additional systematic uncertainty of
$\sigma_{SK}=\plumin{2.9}{2.6}\%$
(excluding $^8$B shape, energy scale and resolution),
which was added to the uncorrelated systematic uncertainties
of the spectrum shape.
\label{tab:zenspec}}
\begin{center}
\footnotesize
\begin{tabular}{|cc|c|ccc|}
\hline
Bin & Range [MeV] & Rate$\pm$stat$\pm$syst [SSM] &
$^8$B Spectrum & E-Scale & E-Resol. \cr
\hline
1 &  5.0-5.5  & 0.4671$\pm$0.0404$\plumi{0.0165}{0.0138}$ &
$\plumi{0.04}{0.02}\%$ & $\plumi{0.09}{0.01}\%$ & $\plumi{0.23}{0.21}\%$ \cr
2 &  5.5-6.5  & 0.4580$\pm$0.0141$\plumi{0.0066}{0.0065}$ &
$\plumi{0.13}{0.09}\%$ & $\plumi{0.20}{0.16}\%$ & $\plumi{0.19}{0.17}\%$ \cr
3 &  6.5-8.0  & 0.4729$\pm$0.0084$\pm{0.0065}$ &
$\plumi{0.41}{0.38}\%$ & $\plumi{0.63}{0.62}\%$ & $\plumi{0.17}{0.16}\%$ \cr
4 &  8.0-9.5  & 0.4599$\pm$0.0093$\pm{0.0063}$ &
$\plumi{0.89}{0.85}\%$ & $\plumi{1.3}{1.3}\%$   & $\plumi{0.12}{0.12}\%$ \cr
5 &  9.5-11.5 & 0.4627$\pm$0.0103$\pm{0.0063}$ &
$\plumi{1.7}{1.6}\%$   & $\plumi{2.5}{2.4}\%$   & $\plumi{0.16}{0.18}\%$ \cr
6 & 11.5-13.5 & 0.4621$\pm$0.0168$\pm{0.0063}$ &
$\plumi{3.1}{2.7}\%$   & $\plumi{4.4}{4.1}\%$   & $\plumi{1.1}{1.1}\%$   \cr
7 & 13.5-16.0 & 0.5666$\pm$0.0390$\pm{0.0078}$ &
$\plumi{5.1}{4.2}\%$   & $\plumi{7.0}{6.4}\%$   & $\plumi{3.2}{3.2}\%$   \cr
8 & 16.0-20.0 & 0.5554$\pm$0.1458$\pm{0.0076}$ &
$\plumi{7.7}{5.6}\%$   & $\plumi{10.6}
{\hspace*{5pt}9.6}\%$ & $\plumi{8.4}{7.9}\%$   \cr
\hline
\multicolumn{2}{|c|}{Comb. 5.0-20.0} &
                0.4653$\pm{0.0047}\plumi{0.0138}{0.0122}$ &
$\plumi{1.15}{1.04}\%$ & $\plumi{1.66}{1.58}\%$ & $\plumi{0.33}{0.34}\%$\cr
\hline
\end{tabular}
\end{center}
\caption{Subdivision of bins 2--7 according to the solar zenith angle
$\theta_z$. The range of $\cos \theta_z$ is given for each bin:
$\cos \theta_z<0$ is `Day' and $\cos \theta_z>0$ is `Night'
(`Mantle' and `Core').
The rates are given in units of 0.001$\times$SSM.
Only statistical uncertainties are quoted. All systematic uncertainties
(see table~\ref{tab:zenspec}) are assumed to be fully correlated in
zenith angle.}
\label{tab:zenspec2}
\begin{center}
\footnotesize
\begin{tabular}{|c|c|ccccc|c|}
\hline
       & Day   & \multicolumn{5}{|c|}{Mantle} & Core \cr
\hspace*{-5pt}Bin\hspace*{-5pt}    & 
-1--0 & 0.00--0.16 & 0.16--0.33 & 0.33--0.50 & 0.50--0.67 & 0.67--0.84 & 0.84--1 \cr
\hline
2 & 453$\pm$20  & 442$\pm$53  & 379$\pm$49  &
    472$\pm$45  & 522$\pm$45  & 503$\pm$49  & 426$\pm$52  \cr
3 & 474$\pm$12  & 530$\pm$34  & 506$\pm$30  &
    438$\pm$26  & 478$\pm$26  & 451$\pm$28  & 439$\pm$31  \cr
4 & 448$\pm$13  & 463$\pm$36  & 470$\pm$33  &
    462$\pm$29  & 509$\pm$29  & 461$\pm$32  & 451$\pm$35  \cr
5 & 453$\pm$15  & 449$\pm$40  & 502$\pm$38  &
    451$\pm$32  & 473$\pm$32  & 477$\pm$35  & 483$\pm$40  \cr
6 & 477$\pm$25  & 509$\pm$67  & 351$\pm$55  &
    391$\pm$49  & 498$\pm$53  & 434$\pm$56  & 521$\pm$64  \cr
7 & 511$\pm$54  & 570$\pm$150 & 831$\pm$167 &
    694$\pm$131 & 665$\pm$127 & 441$\pm$118 & 469$\pm$131 \cr
\hline
\end{tabular}
\end{center}
\end{table}

If combined with the SNO rate\cite{sno} of
$0.346\plumin{0.029}{0.028}\times$SSM
(Appendix, eq.~\ref{eq:sksno}), the SK
rate provides a probe for the appearance of other neutrino
flavors. The SK rate of
$0.465\plumin{0.015}{0.013}\times$SSM
(see table~\ref{tab:zenspec})
exceeds the $e$-type
rate inferred from the SNO measurement by more than $3\sigma$.
If this is interpreted as appearance,
the other flavors contribute about 25\% to the SK rate and
70\% to the $^8$B flux
(the SK cross section for the other flavors
is six to seven times smaller than for $e$-type neutrinos).
The two light-grey allowed regions of figure~\ref{fig:global}a)
are based on the combined fit to the SK zenith angle spectrum and the
SNO rate\cite{venice}. One region contains
the upper part of the LMA solution, the other region
contains the lower part of the LOW solution and the
quasi-VAC solutions (best-fit). The allowed areas
are still independent of the SSM neutrino fluxes.

The Appendix (eq.~\ref{eq:tot}) explains the method to add
the Gallium\cite{gallex,sage} ($74.8\plumin{5.1}{5.0}$SNU or
$0.584\plumin{0.040}{0.039}\times$SSM)
and Chlorine rates\cite{cl} ($2.56\pm0.23$SNU or
$0.337\pm0.030\times$SSM) to the fit.
The allowed areas of this global fit
(shown in figure~\ref{fig:global}b) look quite different
when compared with the $e-$type rate fit:
Only the upper part of the LMA survives (and two tiny quasi-VAC
solutions). The disappearance of the LOW solution
is due to predicted daily variations. The SK rate
requires a larger $^8$B flux than the Chlorine rate allows,
so almost all the quasi-VAC regions disappear.
The fit does not depend on the $^8$B and
{\it hep} neutrino flux predictions by the SSM, which
suffer from the largest uncertainties. However, it depends
on the other SSM neutrino fluxes, in particular the $^7$Be
flux (10\% uncertainty) and the neutrino fluxes of the CNO
cycle ($\approx20\%$ uncertainty). Those fluxes contribute\cite{ssm}
about 15\%($^7$Be) and 6\%(CNO) to the Chlorine rate 
and 27\%($^7$Be) and 7\%(CNO) to the Gallium rate in the SSM.

\section{Results of the global fit}

\begin{table}
\caption{Parameters for the best-fit points. 
The probabilities given in the fourth row are
based on the difference of $\chi^2$
(with respect to the minimum).
The five rows below show the five independent
parts of the fit: the $\Delta\chi^2$ from a
fit to the shape of the SK zenith spectrum
(Appendix, eq.~\protect\ref{eq:sk})
and four
interaction rates (deviation probabilities
are given in units of Gaussian standard deviation
$\sigma$).
The last four rows show the values of the
minimized fit parameters.
The $^8$B and the {\it hep} fluxes are free,
the $^8$B neutrino spectrum shift as well
as the SK energy scale and resolution shifts
are constrained within the systematic uncertainties.} 

\label{tab:solution}
\begin{center}
\footnotesize
\begin{tabular}{|l|cccc|}
\hline
Solution &
Large Mixing&\hspace*{-1mm}Quasi-Vacuum\hspace*{-1mm}&Low $\Delta m^2$& Small Mixing\cr
& Angle (LMA)     &   (Quasi-VAC)      &       (LOW)        & Angle (SMA) \cr
\hline
$\Delta m^2$                    &
6.0$\times10^{-5}$ & 4.57$\times10^{-10}$ & 5.0$\times10^{-8}$ & 4.8$\times10^{-6}$ \cr
$\tan^2\theta$                  &
0.35               & 2.1                  & 0.83               & 0.00044            \cr
\hline
$\chi^2$ (45 dof; $p_{\chi^2}$ [\%]) &
43.4  (53.9)       & 48.5 (33.4)          & 51.2 (24.3)        & 54.2 (16.2)        \cr
$\Delta\chi^2$(2 dof;$p_{\Delta\chi^2}$[\%]) &
 0.0 (100.0)       &  5.1  (7.9)          &  7.8  (2.0)        & 10.8  (0.5)        \cr
\hline
$\Delta\chi^2_{SK}$ ($p_{\Delta \chi^2}$[$\sigma$]) &
 3.4 ( $1.3\sigma$) &  3.1 ( $1.3\sigma$) & 3.9 ( $1.5\sigma$)& 5.0  ( $1.7 \sigma$)        \cr
Ga Rate [SNU]                   & 
73.2 ($-0.3\sigma$)& 69.6  ($-1.0\sigma$) &68.2  ($-1.3\sigma$)&75.1  ($+0.1\sigma$)\cr
Cl Rate [SNU]                   &
2.97 ($+1.8\sigma$)&  3.18 ($+2.8\sigma$) & 3.13 ($+2.5\sigma$)& 2.67 ($+0.5\sigma$)\cr
SK Rate [\%SSM]                 &
46.4 ($-0.1\sigma$)& 44.7  ($-1.4\sigma$) &44.9  ($-1.2\sigma$) & 44.1 ($-1.9\sigma$)\cr
SNO Rate [\%SSM]                &
32.8 ($-0.7\sigma$)& 37.1  ($+0.8\sigma$) &38.5  ($+1.3\sigma$) & 43.8 ($+3.1\sigma$)\cr
\hline
$\phi_{^8B}$ [$10^6/($cm$^2$s)] &
5.62 ($+0.6\sigma$)& 3.71 ($-1.7\sigma$)  & 4.04 ($-1.2\sigma$)& 2.71 ($-2.9\sigma$)\cr
$\phi_{hep}$ [$10^3/($cm$^2$s)] &
40                 & 0                    & 21                 & 8                  \cr
$^8$B Spectrum Shape            &
$-0.3\sigma$       & $-0.7\sigma$         & $-0.1\sigma$       & $+0.1\sigma$       \cr
SK E-scale/resol.               &
$-0.4\sigma$/$-0.1\sigma$       & $-1.0\sigma$/ $0.0\sigma$ & 
$-0.1\sigma$/$-0.2\sigma$       & $+0.1\sigma$/$-0.4\sigma$ \cr
\hline
\end{tabular}
\end{center}
\end{table}

\begin{figure}[p]
\noindent a) \hspace*{13pc} b)
\vspace*{-1.5pc}

\epsfxsize=14.5pc
\epsfbox{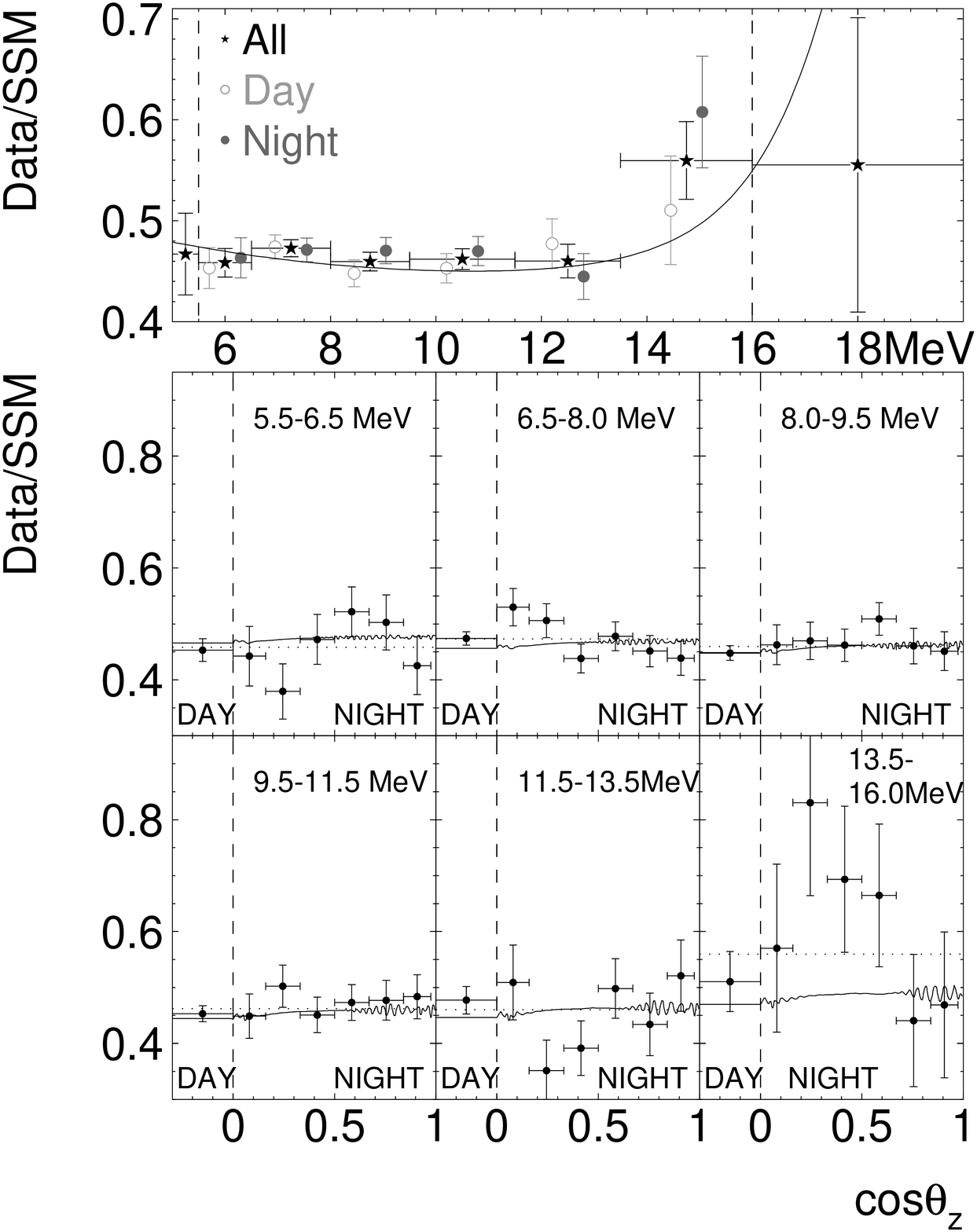} 
\epsfxsize=14.5pc
\epsfbox{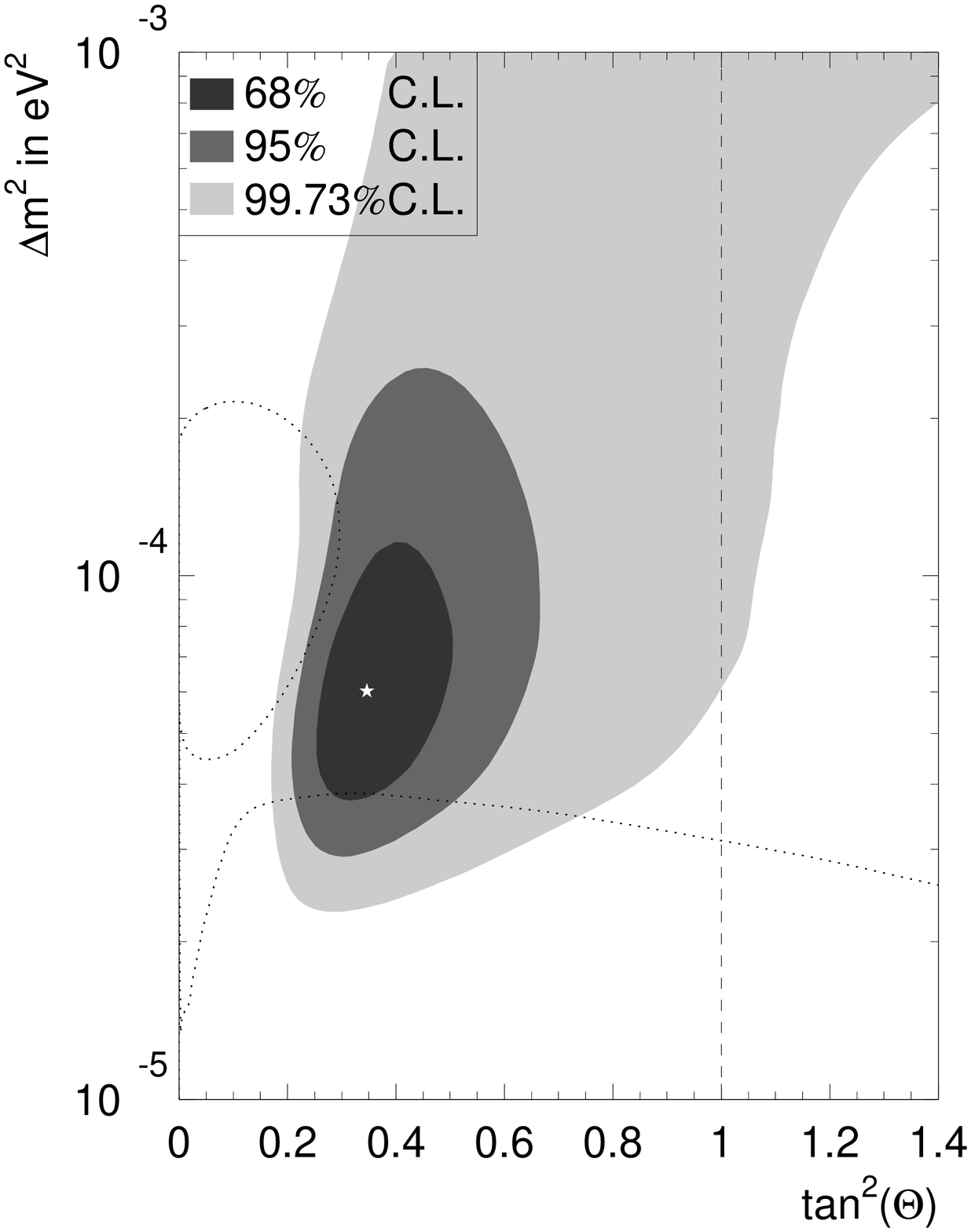} 

\vspace*{-1.5pc}
\noindent c) \hspace*{13pc} d)
\vspace*{-1.5pc}

\epsfxsize=14.5pc
\epsfbox{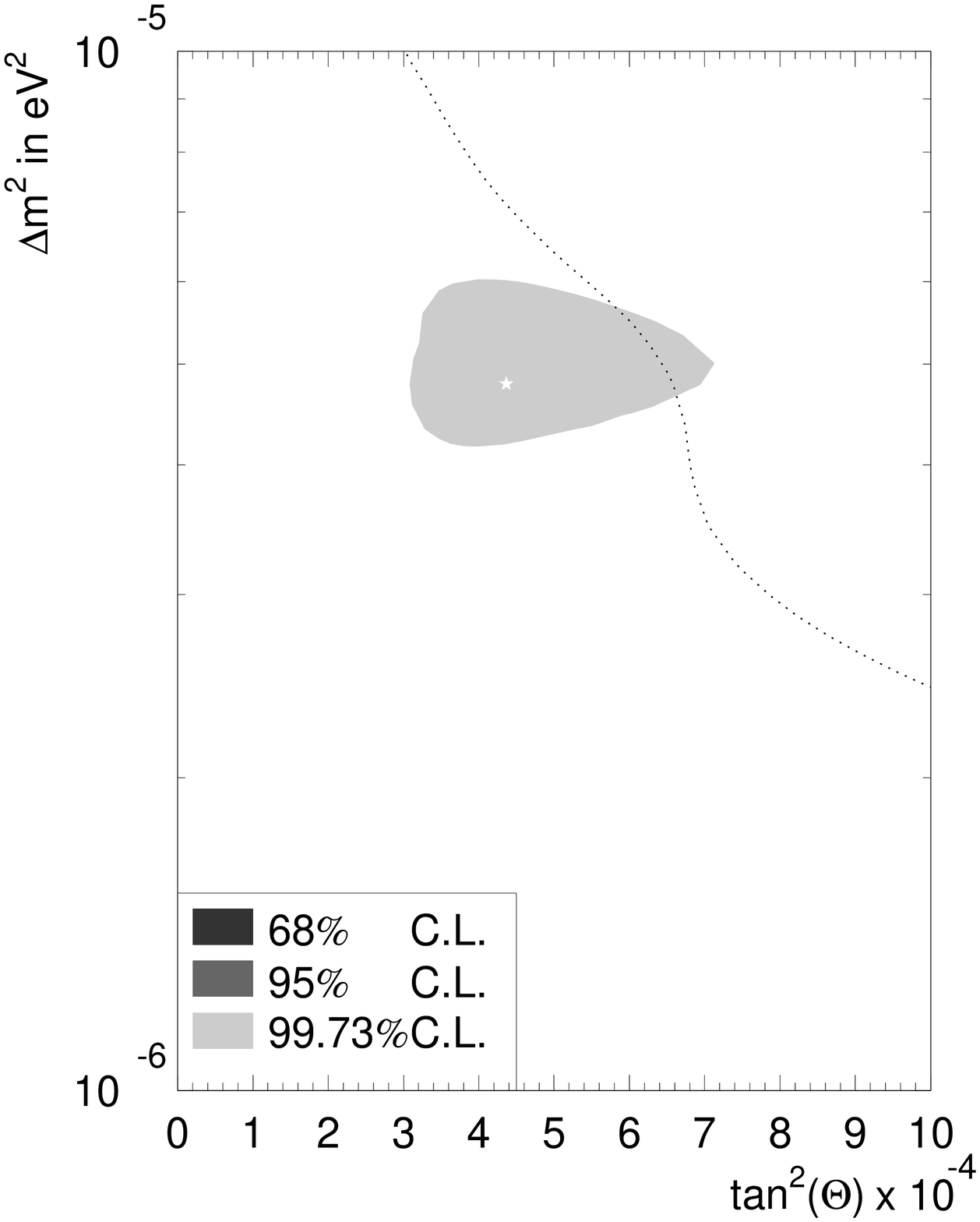} 
\epsfxsize=14.5pc
\epsfbox{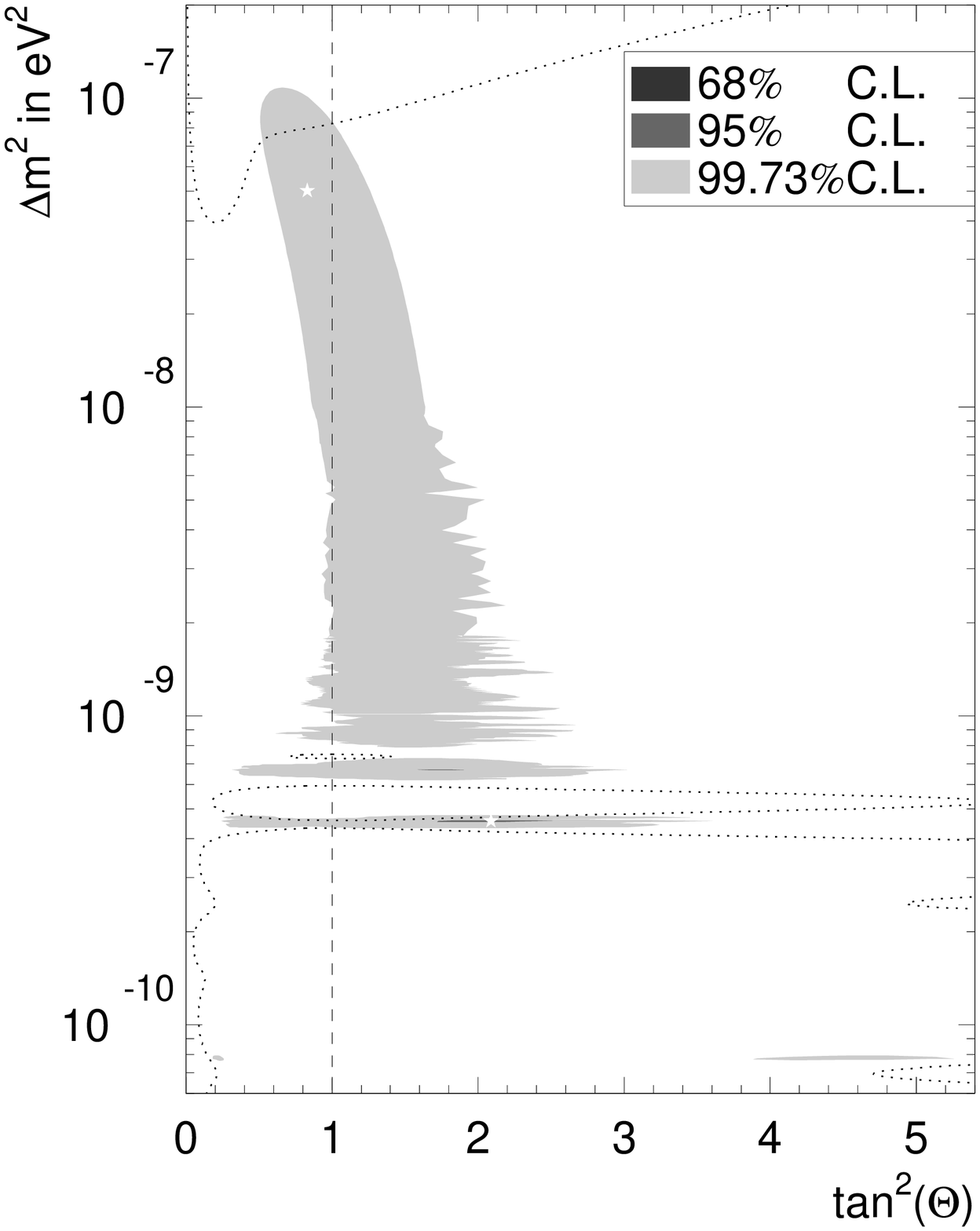} 
\vspace*{-2pc}
\caption{
SK Zenith Spectrum (a) and LMA (b),
SMA (c) and LOW/Quasi-VAC (d) contours
resulting from the global fit.
The top panel of a) shows the SK energy spectrum
during the day (open circles), the night (filled circles)
as well as day and night combined (stars). Between
5.5 and 16 MeV, zenith-angle distributions are shown (lower
6 panels of a). The dotted line in each of the lower panels
is the combined rate in that bin. Error bars reflect the
statistical uncertainty only. Superimposed is an oscillation
prediction ($\tan^2\theta=0.34$
and $\Delta m^2$=6.0$\cdot10^{-5}$eV$^2$) near the best-fit point.
In b)-d) the one (black), two (dark-grey) and three $\sigma$ (light-grey)
contours of the LMA, SMA and LOW/Quasi-VAC solutions are shown
in a linear scale in $\tan^2\theta$.
The best fits are indicated by the white stars. Superimposed
is the 95\%C.L. excluded area from the SK Zenith Spectrum
(inside dotted lines). SMA, LOW and Quasi-VAC are disfavored.}
\label{fig:lma}
\end{figure}

Table~\ref{tab:solution} compares the four smallest local minima of the $\chi^2$
describing the fit. The best-fit is located in the upper LMA area.
The $^8$B flux resulting from this fit is somewhat higher than expected by the SSM
($5.05\plumin{1.01}{0.81}\times10^6/($cm$^2$s)) but well within the uncertainty.
The {\it hep} flux is considerably higher than expected by the SSM
($9.3\times10^3/($cm$^2$s)); however, the uncertainty of this prediction is thought
to be very large.
The fit agrees with the SK zenith angle spectrum moderately well.
Figure~\ref{fig:lma}a) shows the SK zenith angle spectrum and this
best fit. Figure~\ref{fig:lma}b) gives a magnified view of the LMA region (in a linear
scale of $\tan^2\theta$). The $1\sigma$ (black) and $3\sigma$ (light grey)
contours are also shown.
The fit accommodates well the Gallium, SK and SNO rates, however the predicted
Chlorine rate is about $2\sigma$ too high. This worsens the otherwise very
good best-fit $\chi^2$. As a consequence, quasi-VAC solutions, which don't
fit well the interaction rates in general, appear at 95\% C.L. 

\begin{figure}[b]
\epsfxsize=29pc
\epsfbox{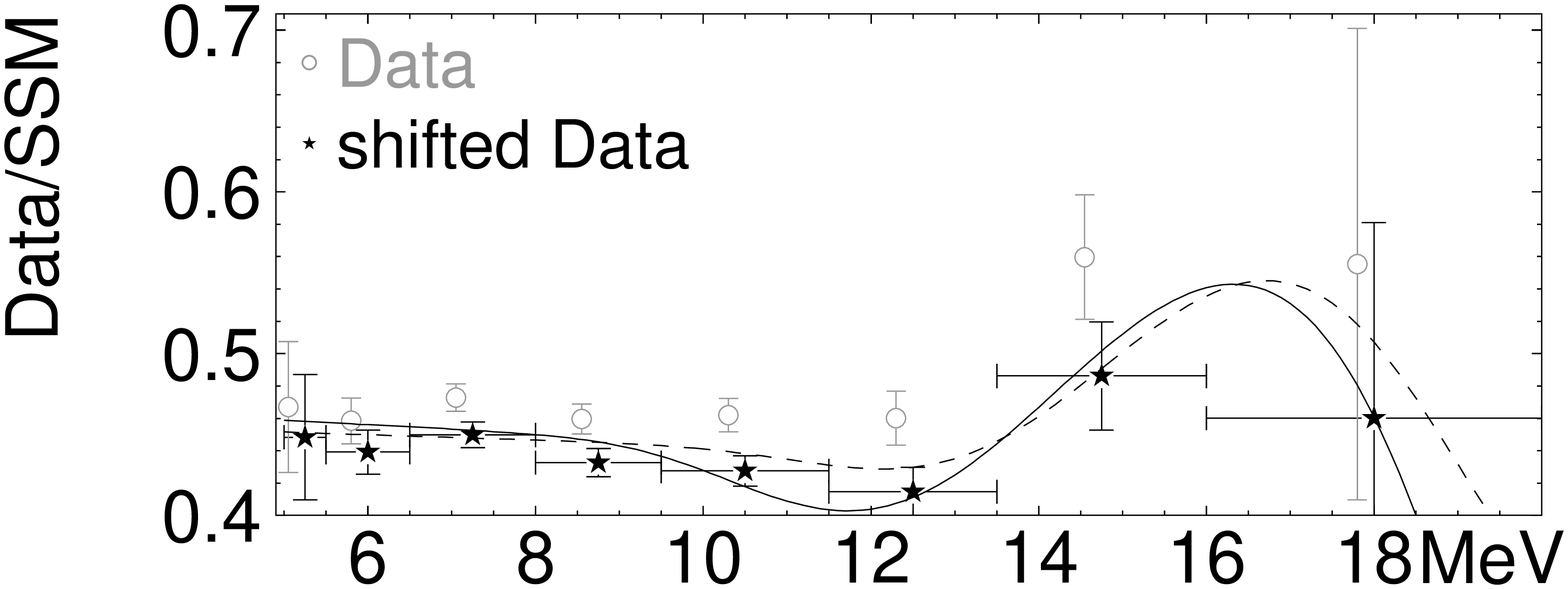} 
\caption{SK Spectral distortion for two quasi-VAC solutions
(solid line: best fit at $4.57\cdot10^{-10}$eV$^2$,
dashed line: $6.68\cdot10^{-10}$eV$^2$)
and data (open circles).
Although the data do not support spectral distortions, a shift (stars)
in energy scale ($-1.0\sigma$) and $^8$B shape ($-0.7\sigma$)
and systematic uncertainty of the combined rate ($-1.4\sigma$) can accommodate the
prediction.}
\label{fig:quasivac}
\end{figure}

The quasi-VAC solution describes the SK zenith angle spectrum slightly
better than the LMA. The fit is helped by a statistical
up-fluctuation of the last two bins (see figure~\ref{fig:quasivac}).
Together with a down-shift of energy scale and $^8$B spectrum, the required
oscillation minimum can be `generated' in the data.
The rates, however, do not fit well. The fit
struggles to accommodate both the Chlorine and SK rate. In this region
Chlorine and SK disagree by $3\sigma$ about the $^8$B flux (which in
the un-oscillated SSM contributes about 76\% of the Chlorine rate).
The resulting 
best-fit $^8$B flux falls $1.7\sigma$ short of the SSM prediction.
Even though the quasi-VAC solution has an overall `C.L. threshold' of 8\%,
it is considerably disfavored, when thus checked in detail.
The surviving (at 98\% C.L.) LOW solution fits the rates about as poorly as
the quasi-VAC solution.
The SK zenith angle spectrum fits somewhat worse than either LMA or quasi-VAC.
The lack of zenith angle variation in the SK data reduces the $\Delta m^2$
(usually around $10^{-7}$eV$^2$) and worsens the LOW best-fit which is already
under pressure from the rates.
Figure~\ref{fig:lma}d) shows an enlarged view of the LOW and quasi-VAC region.

A ``smaller-than-small mixing angle'' solution appears just left of the SMA at about
the $3\sigma$ level (see figure~\ref{fig:lma}c).
The SMA region is defined by the crossing of the Gallium and the Chlorine allowed
area (for a given $^8$B flux). It therefore fits those two rates very well.
The SK zenith angle spectrum, however, fits the worst of all solutions:
The SK spectral data lack the predicted distortion.
The SNO rate is more than $3\sigma$ above the measurement, the SK rate is
too low (2$\sigma$). The $^8$B flux
required is very low ($3\sigma$).

The $^8$B flux comparisons above refer to the SSM value. A recent precision
measurement of the cross section of the $^7$Be(p,$\gamma$)$^8$B fusion reaction
by Junghans et al.\cite{junghans}
implies (see~\cite{garcia})
a flux of $5.93\plumin{0.83}{0.89}\times10^6/($cm$^2$s). If the fit is
confined to this flux, then
all solutions other than LMA are further disfavored.

\section{Conclusion}
A global fit to all solar neutrino data was performed using a two-neutrino
oscillation model. At 95\% C.L. only the upper LMA solution and two tiny
quasi-VAC regions are still allowed. The quasi-VAC regions are disfavored
at more than 92\% C.L. caused by a disagreement about the $^8$B flux
between SK and the Chlorine
rate of $3\sigma$; the resulting quasi-VAC $^8$B flux fit is $1.7\sigma$
smaller than the Standard Solar Model. The LMA solutions above
$\Delta m^2=3\cdot10^{-5}$eV$^2$ are therefore the only viable solution
at 95\% C.L.

\section*{Acknowledgments}
This analysis relies crucially on Super-Kamiokande data.
The author acknowledges the extensive help and cooperation of the 
Super-Kamiokande collaboration as well as the Kamioka Mining and
Smelting Company. The Super-Kamiokande detector has been built and
operated from funding by the Japanese Ministry of Education, Culture,
Sports, Science and Technology, the U.S. Department of Energy, and the
U.S. National Science Foundation.

\section*{Appendix: Fitting Method}
To estimate and limit oscillation parameters, a $\chi^2$ describing
the shape of the SK zenith spectrum is extended to take into account
the measured neutrino interaction rates by various experiments.
Using oscillation probabilities (obtained as in~\cite{osc}) and
SSM neutrino fluxes, the expected
interaction rates $B^{\mbox{\tiny osc}}_{i,z}$ (due to $^8$B neutrinos)
and $H^{\mbox{\tiny osc}}_{i,z}$ (due to {\it hep} neutrinos)
in energy bin $i$ and zenith-angle bin $z$ are calculated. These rates as well
as the SK measurements $D_{i,z}$ are then
normalized by the SSM expectations without oscillations:
\be
b_{i,z}=\frac{B^{\mbox{\tiny osc}}_{i,z}}{B^{\mbox{\tiny SSM}}_{i,z}+H^{\mbox{\tiny SSM}}_{i,z}}
\mbox{ , }
h_{i,z}=\frac{H^{\mbox{\tiny osc}}_{i,z}}{B^{\mbox{\tiny SSM}}_{i,z}+H^{\mbox{\tiny SSM}}_{i,z}}
\mbox{ , }
d_{i,z}=\frac{D_{i,z}}{B^{\mbox{\tiny SSM}}_{i,z}+H^{\mbox{\tiny SSM}}_{i,z}}.
\label{eq:def}
\ee
The zenith components of the flux difference vector $\overrightarrow{\Delta_i}$
\[
\Delta_{i,z}(\beta,\eta)=\left(\beta\cdot b_{i,z} +\eta\cdot h_{i,z}\right)
\times f(E_i,\delta_B,\delta_S,\delta_R)-d_{i,z}
\]
allow for arbitrary neutrino fluxes (through the free parameters $\beta$ and $\eta$).
The combined rate prediction is modified by the energy-shape factors
\[
f(E_i,\delta_B,\delta_S,\delta_R)=
f_B(E_i,\delta_B)\times f_S(E_i,\delta_S)\times f_R(E_i,\delta_R)
\]
with $\delta_B$ describing the $^8$B neutrino spectrum shape uncertainty,
$\delta_S$ describing the uncertainty of the SK energy scale (0.64\%) and
$\delta_R$ describing the uncertainty of the SK energy resolution (2.5\%)
(The shapes $f_B, f_S, f_R$ are given in table~\ref{tab:zenspec}). 
All three uncertainties affect the bins of the SK zenith angle spectrum in a correlated
way. The $7\times 7$ matrices $V_i$ describe statistical and
energy-uncorrelated uncertainties; the latter are assumed to be fully correlated
in zenith angle.
For any given parameters $\delta_k$, the $\chi^2$
\[
\chi^2_{0}=
\sum_{i=1}^{8}
\overrightarrow{\Delta_i}\cdot V_i^{-1}\cdot\overrightarrow{\Delta_i}
=\chi^2_{0,m}+
\left(\overrightarrow{\phi}-\overrightarrow{\phi}_{0,m}\right)C_0
\left(\overrightarrow{\phi}-\overrightarrow{\phi}_{0,m}\right)
\]
\[
\mbox{with}\hspace*{5mm}
C_0=\sum_{i=1}^{8}
\pmatrix{\overrightarrow{b_i}\cdot V_i^{-1}\cdot\overrightarrow{b_i}&
         \overrightarrow{h_i}\cdot V_i^{-1}\cdot\overrightarrow{b_i}\cr
         \overrightarrow{h_i}\cdot V_i^{-1}\cdot\overrightarrow{b_i}&
         \overrightarrow{h_i}\cdot V_i^{-1}\cdot\overrightarrow{h_i}}
\hspace*{5mm}\mbox{ and }\hspace*{5mm}
\overrightarrow{\phi}=\pmatrix{\beta \cr \eta}
\]
can be written as a quadratic form of $\overrightarrow{\phi}$ and the curvature matrix 
$C_0$. The minimum is
\[
\chi^2_{0,m}=\sum_{i=1}^{8}
\overrightarrow{d_i}\cdot V_i^{-1}\cdot\overrightarrow{d_i}-C_{0,m}
\hspace*{5mm}\mbox{with}\hspace*{5mm}
C_{0,m}=\overrightarrow{\phi}_{0,m} C_0
\overrightarrow{\phi}_{0,m}.
\]
If the minimum flux vector is scaled by $\alpha$
($\overrightarrow{\phi}=\alpha\times\overrightarrow{\phi}_{0,m}$)
then $\chi^2_0$ constrains $\alpha$ to be
$\alpha=1\pm\sqrt{1/C_{0,m}}$.
To take into account the systematic uncertainty of the SK combined rate
$\sigma_{SK}=\plumin{2.9}{2.6}\%$
(which is completely correlated in zenith-angle and energy), $\chi^2_0$ is modified
to
\[
\chi^2_1=\chi^2_{0,m}+
\left(\overrightarrow{\phi}-\overrightarrow{\phi}_{0,m}\right)C_1
\left(\overrightarrow{\phi}-\overrightarrow{\phi}_{0,m}\right)
\hspace*{3mm}\mbox{with}\hspace*{3mm}\
C_1=\frac{1/\sigma_{SK}^2}{C_{0,m}+1/\sigma_{SK}^2}\times C_0.
\]
$\chi^2_1$ constrains $\alpha$ to be
$\alpha=1\pm\sqrt{1/C_{0,m}+\sigma_{SK}^2}$, that is, the
minimum is unchanged, but the allowed parameter range for $\beta$
and $\eta$ is larger.
The total $\chi^2$ for the SK zenith spectrum shape is then
\be
\chi^2_{\mbox{\tiny SK}}=
\mbox{Min}\left(\chi^2_1(\beta,\eta,\delta_B,\delta_S,\delta_R)
+\left(\frac{\delta_B}{\sigma_B}\right)^2
+\left(\frac{\delta_S}{\sigma_S}\right)^2
+\left(\frac{\delta_R}{\sigma_R}\right)^2
\right)
\label{eq:sk}
\ee
where all $\delta_k$ as well as $\beta, \eta$ are minimized.

A combined fit with the rate measured by the SNO collaboration\cite{sno}
can be done by defining $b_{\mbox{\tiny SNO}}, h_{\mbox{\tiny SNO}}$ and
$\Delta_{\mbox{\tiny SNO}}(\beta,\eta)$ in a similar way as 
in~(\ref{eq:def})
and form
\be
\chi^2_{\mbox{\tiny SK-SNO}}=
\mbox{Min}\left(\chi^2_1
+\left(\frac{\delta_B}{\sigma_B}\right)^2
+\left(\frac{\delta_S}{\sigma_S}\right)^2
+\left(\frac{\delta_R}{\sigma_R}\right)^2
+\left(\frac{\Delta_{\mbox{\tiny SNO}}(\beta,\eta)}{\sigma_{\mbox{\tiny SNO}}}\right)^2
\right)
\label{eq:sksno}
\ee
To add the radio-chemical rate measurements of Homestake\cite{cl},
Gallex/GNO\cite{gallex}, and SAGE\cite{sage} (all ``Gallium'' rates
are combined into a single rate), the $\chi^2$
\[
\chi^2_{\mbox{\tiny RC}}(\beta,\eta)
\]

takes into account the correlations between Gallium and Chlorine
measurements. The $^8$B and {\it hep} fluxes are constrained by
the minimization of $\chi^2_{\mbox{\tiny SK-SNO}}$, not by the
Standard Solar Model.
The total $\chi^2$ is then a simple addition
\be
\chi^2=\chi^2_{\mbox{\tiny SK-SNO}}+
\mbox{Min}\left(\chi^2_{\mbox{\tiny RC}}
(\alpha\beta_{\mbox{\tiny min}},\alpha\eta_{\mbox{\tiny min}})
+\left(\frac{\alpha-1}{\sigma_\alpha}\right)^2
\right)
\label{eq:tot}
\ee
where
$1/\sigma_\alpha^2=C_{\mbox{\tiny SK-SNO,m}}=\overrightarrow{\phi}_{\mbox{\tiny SK-SNO,m}}
\cdot C_{\mbox{\tiny SK-SNO}}\cdot\overrightarrow{\phi}_{\mbox{\tiny SK-SNO,m}}$.
Numerically, $\sigma_\alpha$ is found to be about 2.7\%
(The accuracy of SK and SNO combined)

\end{document}